# Dynamic Geometry-Based Stochastic Channel Modeling for Polarized MIMO Systems with Moving Scatterers

Hamed Radpour, Laxmikant Minz, Seong-Ook Park, *Senior, IEEE*, Duck-Yong Kim, Young-Chan Moon

*Abstract*— This paper introduces a four-dimensional (4D) geometry-based stochastic model (GBSM) for polarized multiple-input multiple-output (MIMO) systems with moving scatterers. We propose a novel motion path model with high degrees of freedom based on the Brownian Motion (BM) random process for randomly moving scatterers. This model is capable of analyzing the effect of both deterministically and randomly moving scatterers on channel properties. The mixture of Von Mises Fisher (VMF) distribution is considered for scatterers resulting in a more general and practical model. The proposed motion path model is applied to the clusters of scatterers with the mixture of VMF distribution, and a closed form formula for calculating space time correlation function (STCF) is achieved, allowing the study of the behavior of channel correlation and channel capacity in the time domain with the presence of stationary and moving scatterers. To obtain numerical results for channel capacity, we employed Monte Carlo simulation method for channel realization purpose. The impact of moving scatterers on the performance of polarized MIMO systems is evaluated using $2\times 2$ MIMO configurations with various dual polarizations, i.e. V/V, V/H, and slanted $\pm 45°$ polarizations for different signal-to-noise (SNR) regimes. The proposed motion path model can be applied to study various dynamic systems with moving objects. The presented process and achieved formula are general and can be applied to polarized MIMO systems with any arbitrary number of antennas and polarizations.

## I. INTRODUCTION

MODERN wireless communication systems are swiftly moving towards tremendous data transfer due to growing demands for emerging dynamic applications of Internet of Things (IoT), Vehicle-to-Vehicle (V2V), Air-to-Air (A2A) and Air-to-Ground (A2G) communications such as unmanned aerial vehicle-ground (UAV-ground) systems. Among the dynamic A2G communications, unmanned aerial vehicle-intelligent reconfigurable surface (UAV-IRS) systems [1-3] have gained a great attention in the last two years due to their key role in providing very high capacity and low latency for sixth-generation (6G) communication systems. These applications along with the next generation communication systems necessitate developed nonstationary channel models to encompass new aspects of modern systems.

In the last two decades, MIMO channel models have been continuously evolved to not only consider emerging scenarios but also study them in a more detailed and practical form. The more complete and convenient to use a channel model would be, the more worthwhile and applicable it will be. Among the proposed 3D channel models, [4-5] have studied the polarized MIMO systems assuming stationary clusters of scatterers. Furthermore, the channel models for MIMO systems can be classified into deterministic [6-10] and stochastic [11-17] models. Among them, deterministic approach takes ray-tracing (RT) or finite-difference-time-domain (FDTD) techniques to obtain channel characteristics. However, these numerical methods have intrinsic disadvantage of requiring a comprehensive and detailed information of the intended propagation environment as a digital map, that makes them to be time-consuming, site-dependent and highly complicated. On the other hand, the stochastic models can be divided into two main categories of non-geometry-based stochastic models (NGBSMs) [18-21] and GBSMs [22-25]. The GBSMs deal with the effect of scatterers in a propagation environment by mapping the effective scatterers on a geometrical model. On the other hand, the NGBSMs specify locations of scatterers based on some random distributions. Although the NGBSMs can study the effect of channel components on the channel characteristics, they are incapable of providing a proper physical insight of the system thus they are unable to be extended to different scenarios. Therefore, the GBSMs which consider the scatterers distribution in a geometrical framework can construct a conceptual and flexible approach.

In this connection, empirical studies [26-29] represent that the stationary assumption is only valid for a trivial time interval. Accordingly, a nonstationary GBSM for V2V with fixed scatterers considering a two-sphere with an elliptic-cylinder channel model is developed in [30]. A nonstationary channel for UAV-MIMO systems utilizing elliptic-cylinder geometry has been studied in [31]. Moreover, a one-ring GBSM with randomly moving mobile station (MS) is provided in [32-34]. These 2-D models consider BM random process to specify MS motion with stationary scatterers distributed on a ring. Therefore, the nonstationarity of channels in these models originates from the motion of transceivers rather than scatterers.

Hamed Radpour, Laxmikant Minz and Seong Ook Park are with the Microwave and Antenna Lab, Department of Electrical Engineering, KAIST, Daejeon, South Korea, 34141. Email (hradpour@kaist.ac.kr)

Duck-Yong Kim and Young-Chan Moon are with KMW Inc., Hwaseong, South Korea, 18642.

Nevertheless, the number of dynamic channel models that investigated the unique aspects of MIMO systems with both moving MS and specially moving scatterers are very few in the literature. A nonstationary single-input single-output (SISO) model considering some random distributions for the velocity of the scatterers is discussed in [35-36]. Even though this model does not provide a proper physical insight, it can be used to study the effect of moving scatterers in special scenarios such as moving vehicles along the roads and highways. A two-ring and ellipse nonstationary GBSM for MIMO V2V communications assuming moving scatterers on the two-rings and stationary scatterers on the ellipse is proposed in [37]. Impact of moving scatterers for a MIMO V2V system is analyzed in [38] considering a two-cylinder model where stationary scatterers are located on the surface of the two cylinders and moving scatterers are distributed on the base of the cylinders and antenna arrays are also positioned at the center of the corresponding cylinder bases. All nonstationary models discussed so far considered omni-directional array antennas while the channel model proposed in [39] took cross-polarized antennas into account and attempted to extend the model of [38] by assuming 2D multi-rings with the same height of the antenna arrays for the moving scatterers. Furthermore, a nonstationary multi-layer cylinder model for UAV-MIMO channels assuming omni-directional antennas is introduced in [40] where local and far clusters are separated and distributed on different cylinders by locating the moving scattereres on the base of these cylinders similar to [37]. Based on the above description, one drawback of the all mentioned models for MIMO channels with moving scatterers in [37-40] is that they consider a 2D model for both location and movement direction of moving scatterers. Another disadvantage is that these models consider the transmitter or receiver to have the same altitude as moving scatterers and this assumption is not accurate for most scenarios even for V2V communications such as at urban areas where the transmitter or receiver antenna arrays are often located in different altitude of surrounding scatterers.

Motivated by the above gaps in the literature of channel modeling, this paper introduces a 4D GBSM for polarized MIMO systems in the presence of randomly moving scatterers. The principle contributions and novelties of the current research are summarized as follows:

- A novel motion path model based on the BM random process is proposed which allows studying of randomly moving objects in various dynamic systems.
- Based on the flexibility of the proposed motion model, moving scatterers are considered to be distributed in a 3D environment thus the direction of the motion is also 3D and arbitrary.
- The effect of both deterministically and randomly moving scatterers on the important channel statistics such as space-time correlation function and channel capacity is analyzed.
- Finally, this paper investigates the impact of moving scatterers in a cross-polarized MIMO system rather than assuming omni-directional antenna elements.

It is worthwhile to mention that the polarized MIMO systems are popular in practice as they occupy less space and provide more channel capacity. Therefore, the characteristics of antenna arrays such as field pattern, polarization and antenna orientation as well as system statistics such as cross-polarization-discrimination (XPD) are contributed to the presented approach. Based on the generality of the proposed model, it can be utilized to study the effect of moving scatterers in different dynamic systems including V2V, A2G and A2A communications.

The remainder of the paper is organized as follows. In Section II, the configuration of a two-spherical nonstationary MIMO channel model and distribution of scatterers is described. Section III introduces a novel Motion Path model based on the Brownian Motion random process. Section IV provides the channel statistics for the developed dynamic model by employing the introduced Motion Path model in Section III. Section V discusses the channel properties of the presented dynamic model. Finally, the conclusions are drawn in Section VI.

## II. CONFIGURATION OF THE DYNAMIC CHANNEL MODEL

To specify the geometrical configuration, consider a MIMO system containing $S$ Uniform Linear Array (ULA) antennas ($Tx_1$, ..., $Tx_S$) on the transmitter side and $U$ ULA antennas ($Rx_1$, ..., $Rx_U$) on the receiver side, as shown in Fig. 1. Now consider a two-spherical model with scatterers located on the surface of spheres moving radially towards or far away from the center of their corresponding array antenna. It is assumed that the receiver is moving with the velocity of **v** and the transmitter is fixed. Suppose that there are $K$ and $L$ scatterers at the transmit and receive side represented by TS$_k$ ($k=1,2,...,K$) and RS$_l$ ($l=1,2,...,L$), respectively. The symbols ($\vartheta_k, \varphi_k$) are denoting the $k$th transmit scatterer angular location on the surface of a sphere with radius of $R^{Tx}(t)$. Likewise, the angles ($\theta_l, \phi_l$) are specifying the $l$th scatterer on the sphere of radius $R^{Rx}(t)$. Therefore, the parameters ($\vartheta_k, \varphi_k$) and ($\theta_l, \phi_l$) are also defining the angle of departure (AoD) and angle of arrival (AoA), respectively. The instantaneous radius of scatterers, $R^{Tx}(t)$ and $R^{Rx}(t)$ can be written as

$$R^{Tx}(t) = R_0^{Tx} + v^{Tx}t \ , \ \ R^{Rx}(t) = R_0^{Rx} + v^{Rx}t \quad (1)$$

where $R_0^{Tx}$ and $R_0^{Rx}$ are indicating the initial radius, $v^{Tx}$ and $v^{Rx}$ are the velocity of clusters at transmitter and receiver side, respectively and they determine whether the scatterers are moving away from their corresponding array center or approaching to it. In fact, $v^{Tx}$ and $v^{Rx}$ are giving further degrees of freedom to analyze the effect of radially moving scatterers and for the case of $v^{Tx} = v^{Rx} = 0$ we just have the Brownian motion which will be explained in the next section. The antenna element spacings at transmitter and receiver are denoted by $\delta_T$ and $\delta_R$, respectively and $(\vartheta^{Tx}, \varphi^{Tx})$ and $(\theta^{Rx}, \phi^{Rx})$ are specifying the orientation of transmit and receive antenna arrays, respectively. The parameter $D$ shows the distance between the center of the transmitter and receiver and $D_{Sk}$ denotes the distance between the $k^{th}$ transmit scatterer and $S^{th}$ transmit antenna element. Similarly, $D_{Ul}$ shows the distance between the $l^{th}$ receive scatterer and $U^{th}$ receive antenna element; $D_{kl}$ is the distance between $k$th and $l$th scatterers.

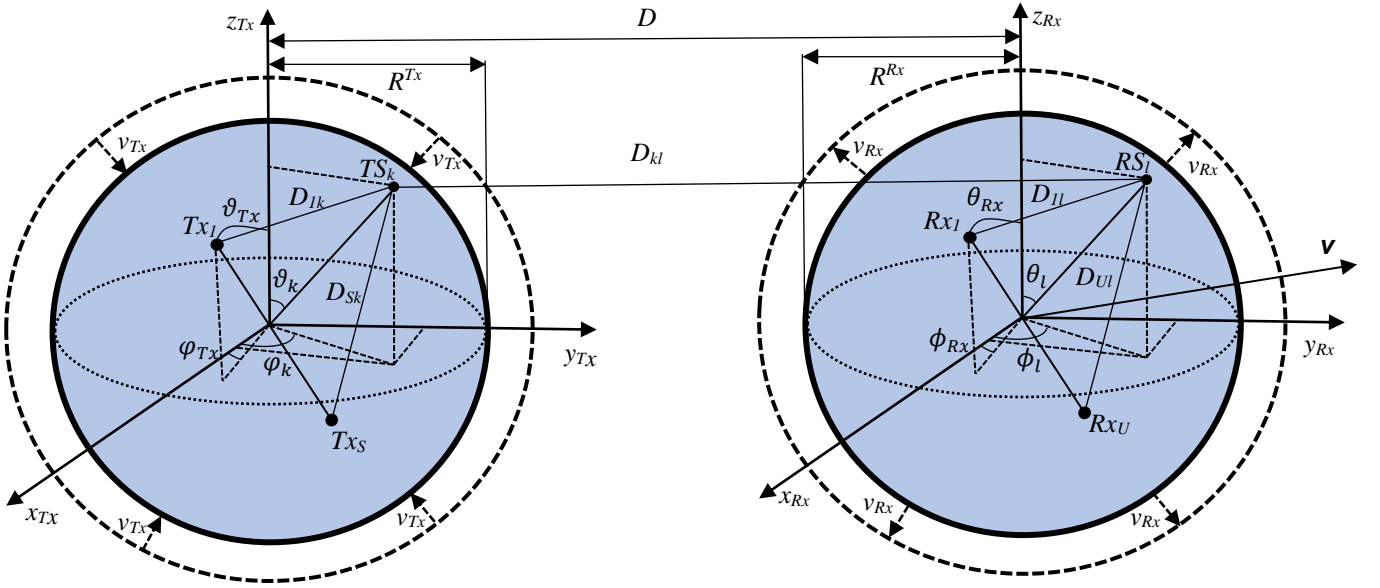

Fig.1. Geometrical channel model with moving scatterers

Scatterers are considered to have the mixture of Von Mises Fisher (VMF) distribution and clusters are assumed to have an identical radial velocity so at all moments every cluster in the transmitter and receiver part will be located on the spherical surface of radius $R^{Tx}(t)$ and $R^{Rx}(t)$, respectively.

The VMF distribution [41] used to model the scatterers in a single cluster is defined by

$$f(\vartheta, \varphi | \vartheta_0, \varphi_0, \kappa) = C_p(\kappa) \exp^{\kappa[\sin\vartheta_0 \sin\vartheta \cos(\varphi - \varphi_0) + \cos\vartheta_0 \cos\vartheta]} \sin\vartheta \quad (2)$$

where $\vartheta_0$ and $\varphi_0$ are the mean azimuth and elevation angles of the cluster, $\kappa$ is the concentration parameter which determines the spread of scatterers around the mean direction; $C_p(\kappa)$ is the normalizing constant as $C_p(\kappa) = \kappa / 4\pi \sinh\kappa$ in the Euclidean space $\mathbb{R}^3$ where $p = 3$. Based on the measurements, a propagation environment is generally distinguished to follow a multiple cluster model of scatterers. Hence, the distribution of a mixture model encompassing Q VMFs can be specified as

$$p(\vartheta, \varphi) = \sum_{q=1}^{Q} \rho_q f_q(\vartheta, \varphi | \vartheta_q, \varphi_q, \kappa_q) \quad (3)$$

where $\rho_q$ is the prior probability of the $q^{th}$ cluster; $\vartheta_q, \varphi_q$ are specifying the mean direction and $\kappa_q$ is the concentration parameter of the $q^{th}$ cluster. Therefore, the distribution of scatterers in both transmitter and receiver side can be represented in a similar way by

$$p^{Tx}(\Omega^{Tx}) = p^{Tx}(\vartheta, \varphi) = \sum_{q=1}^{Q_{Tx}} \rho_q^{Tx} f_q(\vartheta^{Tx}, \varphi^{Tx} | \vartheta_q^{Tx}, \varphi_q^{Tx}, \kappa_q^{Tx})$$

$$p^{Rx}(\Omega^{Rx}) = p^{Rx}(\theta, \phi) = \sum_{q=1}^{Q_{Rx}} \rho_q^{Rx} f_q(\theta^{Rx}, \phi^{Rx} | \theta_q^{Rx}, \phi_q^{Rx}, \kappa_q^{Rx})$$

(4)

where $\Omega^{Tx}$ and $\Omega^{Rx}$ are the solid angles that characterize the scatterer directions for transmitter and receiver, respectively.

### III. Proposed Motion Path Model

In this section, firstly a short background of the Brownian Motion following with the mathematical definition of this random random process will be discussed. Finally, we will introduce the proposed BM path model which is specially suitable to model the moving objects that are mapped and distributed on the surface of a sphere.

In 1828, the botanist scientist, *Robert Brownian* described his observations on the chaotic motion of suspended particles in a fluid or gas. Since his fundamental contribution led to a new discovery to explanation of some significant physical phenomena, the associated theory is called *Brownian Motion*. Henceforth, several scientists attempted to express a formula for this random motion and finally, *Einstein* (1905) accomplished to a provide a mathematical formula [42] that delivers a physical explanation to these phenomena. Afterward, this new branch of stochastic studies became the basis of plenty of complicated processes. Further historical background as well as mathematical approaches are discussed in [43-46].

A stochastic process $\{B(t): t \in [0,T]\}$ is considered a standard BM process if it satisfies the following three conditions:

1) $B(0) = 0$.
2) $\forall 0 \leq u < v \leq T$, the random variable given by the increment $B(v) - B(u)$ follows a Gaussian distribution with zero mean and variance $v - u$ that is $B(v) - B(u) \sim N(0, v-u)$.
3) $B(t)$ has independent increments as $\forall 0 \leq u < v < w < x \leq T$, $B(v) - B(u)$ and $B(x) - B(w)$ are statistically independent.

Therefore, $B(t)$ is a Wiener process with normally distributed and independent increments.

Now, the goal is to propose a motion path model that starts from a starting angular point $(\theta_s, \varphi_s)$ on the considered surface of a sphere and terminates at the destination point $(\theta_d, \varphi_d)$ where the parameters $\theta$ and $\varphi$ are the azimuth and elevation components of spherical coordinate systems, respectively. The trajectory model from the starting point to the destination point consists of both deterministic and random motions. To characterize the deterministic component, we define drift increments as $\delta_\theta = (\theta_d - \theta_s)/M$ and $\delta_\varphi = (\varphi_d - \varphi_s)/M$ where $M$ is a positive integer that determines the number of considered path segmentations.

Now we combine the both deterministic drift component and random component to introduce the *Motion path* $M_p$ model as follows

$$M_p = \begin{cases} \theta(t_m) = \theta_s + \omega_\theta^d m \delta_\theta + \sigma_\theta B_\theta(t_m) \\ \phi(t_m) = \phi_s + \omega_\phi^d m \delta_\phi + \sigma_\phi B_\phi(t_m) \end{cases} \quad (5)$$

Where $m$ is the counting variable of BM as $m \in [0, M]$ and is used to indicate the time slot $t_m$ and the parameters $\omega_\theta^d$ and $\omega_\varphi^d$ are the elevation and azimuth angular speeds in $rad/\sec$ that construct the drift components $\omega_\theta^d t_m \delta_\theta$ and $\omega_\varphi^d t_m \delta_\varphi$ which specify the translational motion. The parameters $\sigma_\theta$ and $\sigma_\varphi$ are the elevation and azimuth standard deviations in $rad$ that map the random Brownian Motion into the $\theta - axis$ and $\phi - axis$, respectively and add to the degree of freedom by controlling the randomness of the *Motion Path*.

It is worthwhile to mention that tracking of direction of moving objects that are distributed on a unit sphere is discussed in [47,48] using Bayesian estimation. However, here we will employ our discussed Brownian Motion model to analyze a dynamic wireless communication system.

IV. Statistical Properties of the Dynamic Channel

In this section the above-mentioned proposed Brownian Motion path model is applied to the predefined 2-spherical model to investigate the nonstationary channel statistics. Thus, assuming scatterers with the mixture of VMF distributions, we consider that the $q^{th}$ cluster is randomly moving from the angular position $(\theta_s, \varphi_s)$ to $(\theta_d, \varphi_d)$ following the described *Motion Path* model. Then the corresponding dynamic distribution of the q$^{th}$ cluster follows the time-variant VMF (TV-VMF) PDF as

$$f(\theta, \varphi | \theta_0(t_m), \varphi_0(t_m), \kappa) = C_p(\kappa)$$
$$\times \exp^{\kappa[\sin\theta_0(t_m)\sin\theta\cos(\varphi - \varphi_0(t_m)) + \cos\theta_0(t_m)\cos\theta]} \sin\theta \quad (6)$$

where the whole cluster moves by motion of the mean angular values $\theta_0(t_m), \varphi_0(t_m)$.

Now a $2 \times 2$ MIMO system with half a wavelength dipole antenna elements is assumed. Then considering $F^{(v)}(\theta', \phi')$ and $F^{(h)}(\theta', \phi')$ as the field patterns for vertical and horizontal polarizations, respectively at point $P(r', \theta', \phi')$ it can be written that [19]

$$F^{(v)}(\theta', \phi') = \left| (\cos\theta'\cos\phi'\sin\gamma - \sin\theta'\cos\gamma) \frac{\cos(\pi\xi/2)}{1-\xi^2} \right|$$
$$F^{(h)}(\theta', \phi') = \left| \sin\phi'\sin\gamma \frac{\cos(\pi\xi/2)}{1-\xi^2} \right| \quad (7)$$

where $\xi = \sin\theta'\cos\phi'\sin\gamma + \cos\theta'\cos\gamma$ and $\gamma$ is the angle between dipole antenna and z axis.

Considering a non-line of sight and flat fading system, the output signal can be represented as

$$y(t) = \mathbf{H}(t)x(t) + n(t) \quad (8)$$

where y(t), x(t) and n(t) are received vector, transmitted vector and noise vector, respectively. H(t) is a $U \times S$ complex channel matrix which can be written as

$$\mathbf{H} = \begin{pmatrix} h_{11}^{norm}(t) & \cdots & h_{1S}^{norm}(t) \\ \vdots & \ddots & \vdots \\ h_{U1}^{norm}(t) & \cdots & h_{US}^{norm}(t) \end{pmatrix} \quad (9)$$

where normalized channel coefficient is denoted by $h_{u,s}^{norm}(t)$ and can be defined as

$$h_{u,s}^{norm}(t) = \frac{h_{u,s}(t)}{\sqrt{E\left[|h_{u,s}(t)|^2\right]}} \quad (10)$$

The spatial correlation between $h_{p,m}^{norm}(t)$ and $h_{q,n}^{norm}(t)$ can be written as

$$r_{pm,qn}(t) = E\left[h_{p,m}^{norm}(t) h_{q,n}^{norm*}(t)\right] = \frac{E\left[h_{p,m}(t) h_{q,n}^{*}(t)\right]}{\sqrt{E\left[|h_{p,m}(t)|^2\right]}\sqrt{E\left[|h_{q,n}(t)|^2\right]}} \quad (11)$$

The spatial correlation matrix $\mathbf{R} = [r_{i,j}]_{US \times US}$ for channel coefficients can be calculated by

$$\mathbf{R} = \text{cov}(\text{vec}(\mathbf{H})) = E\left[\text{vec}(\mathbf{H}) \text{vec}^H(\mathbf{H})\right] \quad (12)$$

Since $\mathbf{R}$ is a Hermitian matrix and $r_{i,i}$ is unity, the spatial correlation matrix behavior can be evaluated by

$$\bar{r} = \frac{\sum_{i<j} r_{i,j}}{US(US-1)/2} \quad (13)$$

Two important parameters to characterize a multi-polarized MIMO system are XPD and co-polar-ratio (CPR). The XPD specifies the leakage rate of a polarization into another one in a propagation environment which quantifies the depolarization. However, the CPR indicates the link quality of a polarization compared with the other one and it can be written as

$$\text{XPD}_h = E\left[|h^{hh}|^2\right] / E\left[|h^{hv}|^2\right]$$
$$\text{XPD}_v = E\left[|h^{vv}|^2\right] / E\left[|h^{vh}|^2\right] \quad (14)$$
$$\text{CPR} = E\left[|h^{vv}|^2\right] / E\left[|h^{hh}|^2\right]$$

where $h^{gf}$ is the $gf$ channel component.

Since the amount of scatterers is considerably large ($K, L \to \infty$) then the numerator of (11) can be written as [4]

$$E[h_{p,m}(t) h_{q,n}^{*}(t)] = \sqrt{P_{p,m}(t) P_{q,n}(t)}$$
$$\iint_{\Omega^{Tx}} \iint_{\Omega^{Rx}} \exp(-jk_0(D_m^{Tx} + D_p^{Rx} - D_n^{Tx} - D_q^{Rx}))$$
$$\times F_m^{Tx}(\Omega^{Tx}) F_p^{Rx}(\Omega^{Rx}) F_n^{Tx}(\Omega^{Tx}) F_q^{Rx}(\Omega^{Rx})$$
$$+ E[1/\text{XPD}_v] F_m^{Tx}(\Omega^{Tx}) F_p^{Rx}(\Omega^{Rx}) F_n^{Tx}(\Omega^{Tx}) F_q^{Rx}(\Omega^{Rx})$$
$$+ E[1/\text{XPD}_h] F_m^T x(\Omega^{Tx}) F_p^{Rx}(\Omega^{Rx}) F_n^{Tx}(\Omega^{Tx}) F_q^{Rx}(\Omega^{Rx})$$
$$+ E[1/\text{CPR}] F_m^{Tx}(\Omega^{Tx}) F_p^{Rx}(\Omega^{Rx}) F_n^{Tx}(\Omega^{Tx}) F_q^{Rx}(\Omega^{Rx}) \quad (15)$$
$$\times p^{Tx}(\vartheta, \varphi) p^{Rx}(\theta, \phi) d\Omega^{Tx} d\Omega^{Rx}$$

where $k_0 = 2\pi / \lambda$ and $\lambda$ is the wavelength.

Assuming $\vec{r}_m^{Tx}$ as the $m$th transmit antenna position vector and $\vec{R}_{\Omega^{Tx}}$ as the position vector associated with the point on the transmit scatterer sphere with the solid angle of $\Omega^{Tx}$, the distance parameter $D_m^{Tx}$ and the term $\cos(\alpha_m)$ are defined as

$$D_m^{Tx} = |\vec{R}_{\Omega^{Tx}} - \vec{r}_m^{Tx}|$$
$$\cos(\alpha_m) = \frac{\vec{R}_{\Omega^{Tx}} \cdot \vec{r}_m^{Tx}}{|\vec{R}_{\Omega^{Tx}} \cdot \vec{r}_m^{Tx}|} \quad (16)$$

The other $D$ parameters can be defined similarly

$$D_m^{Tx}(t) = [R^{Tx}(t)^2 + r_m^2 - 2R^{Tx}(t) r_m \cos(\alpha_m)]^{1/2}$$
$$D_n^{Tx}(t) = [R^{Tx}(t)^2 + r_n^2 - 2R^{Tx}(t) r_n \cos(\alpha_n)]^{1/2}$$
$$D_p^{Rx}(t) = [R^{Rx}(t)^2 + r_p^2 - 2R^{Rx}(t) r_p \cos(\alpha_p)]^{1/2}$$
$$D_q^{Rx}(t) = [R^{Rx}(t)^2 + r_q^2 - 2R^{Rx}(t) r_q \cos(\alpha_q)]^{1/2} \quad (17)$$

Assuming $r_m \ll R^{Tx}$ and by applying the approximation

$$(1+x)^r \approx 1 + rx + \frac{r(r-1)}{2} x^2 \quad (18)$$

the distance parameters become as follow

$$D_m^{Tx} = R^{Tx}(t) - r_m \cos\alpha_m - \frac{r_m^2 \cos^2\alpha_m}{2R^{Tx}(t)}$$
$$D_n^{Tx} = R^{Tx}(t) - r_n \cos\alpha_n - \frac{r_n^2 \cos^2\alpha_n}{2R^{Tx}(t)}$$
$$D_p^{Rx} = R^{Rx}(t) - r_p \cos\alpha_p - \frac{r_p^2 \cos^2\alpha_p}{2R^{Rx}(t)}$$
$$D_q^{Rx} = R^{Rx}(t) - r_q \cos\alpha_q - \frac{r_q^2 \cos^2\alpha_q}{2R^{Rx}(t)} \quad (19)$$

where

$$\cos\alpha_m = \sin\vartheta_m^{Tx} \cos\varphi_m^{Tx} \sin\vartheta \sin\varphi + \sin\vartheta_m^{Tx} \sin\varphi_m^{Tx} \sin\vartheta \sin\varphi + \cos\vartheta_m^{Tx} \cos\vartheta$$

$$\cos\alpha_n = \sin\vartheta_n^{Tx} \cos\varphi_n^{Tx} \sin\vartheta \sin\varphi + \sin\vartheta_n^{Tx} \sin\varphi_n^{Tx} \sin\vartheta \sin\varphi + \cos\vartheta_n^{Tx} \cos\vartheta$$

$$\cos\alpha_p = \sin\theta_p^{Rx} \cos\phi_p^{Rx} \sin\theta \sin\phi + \sin\theta_p^{Rx} \sin\phi_p^{Rx} \sin\theta \sin\phi + \cos\theta_p^{Rx} \cos\phi$$

$$\cos\alpha_q = \sin\theta_q^{Rx} \cos\phi_q^{Rx} \sin\theta \sin\phi + \sin\theta_q^{Rx} \sin\phi_q^{Rx} \sin\theta \sin\phi + \cos\theta_q^{Rx} \cos\phi \quad (20)$$

By defining $D_{mn}(t)$ and $D_{pq}(t)$ as

$$D_{mn}(t) = D_n^{Tx} - D_m^{Tx}$$
$$D_{pq}(t) = D_q^{Tx} - D_p^{Tx} \quad (21)$$

It can be written that

$$D_{mn}(t) = (r_n cos\alpha_n - r_m cos\alpha_m)\left[1 + \frac{1}{2R_T(t)}(r_n cos\alpha_n + r_m cos\alpha_m)\right]$$

$$D_{pq}(t) = (r_q cos\alpha_q - r_p cos\alpha_p)\left[1 + \frac{1}{2R_R(t)}(r_q cos\alpha_q + r_p cos\alpha_p)\right]$$

(22)

Finally, the correlation at (15) can be written as

$$\begin{aligned}
E\left[h_{p,m}(t)h_{q,n}^*(t)\right] &= \sqrt{P_{pm}P_{qn}} \\
&\times (\int_0^{2\pi}\int_0^{\pi} \exp(-jD_{mn}(t)) \\
&\times F_m^{Tx(v)}(\vartheta,\varphi) F_n^{Tx(v)}(\vartheta,\varphi) p^{Tx}(\vartheta,\varphi) sin\vartheta d\vartheta d\varphi \\
&\times \int_0^{2\pi}\int_0^{\pi} \exp(-jD_{pq}(t)) \\
&\times F_p^{Rx(v)}(\theta,\phi) F_q^{Rx(v)}(\theta,\phi) p^{Rx}(\theta,\phi) sin\theta d\theta d\phi \\
&+ E\left[\frac{1}{XPD_v}\right] \times \int_0^{2\pi}\int_0^{\pi} \exp(-jD_{mn}(t)) \\
&\times F_m^{Tx(v)}(\vartheta,\varphi) F_n^{Tx(v)}(\vartheta,\varphi) p^{Tx}(\vartheta,\varphi) sin\vartheta d\vartheta d\varphi \\
&\times \int_0^{2\pi}\int_0^{\pi} \exp(-jD_{pq}(t)) \\
&\times F_p^{Rx(v)}(\theta,\phi) F_q^{Rx(v)}(\theta,\phi) p^{Rx}(\theta,\phi) sin\theta d\theta d\phi \\
&+ E\left[\frac{1}{XPD_h}\right] E\left[\frac{1}{CPR}\right] \int_0^{2\pi}\int_0^{\pi} \exp(-jD_{mn}(t)) \\
&\times F_m^{Tx(v)}(\vartheta,\varphi) F_n^{Tx(v)}(\vartheta,\varphi) p^{Tx}(\vartheta,\varphi) sin\vartheta d\vartheta d\varphi \\
&\times \int_0^{2\pi}\int_0^{\pi} \exp(-jD_{pq}(t)) \\
&\times F_p^{Rx(v)}(\theta,\phi) F_q^{Rx(v)}(\theta,\phi) p^{Rx}(\theta,\phi) sin\theta d\theta d\phi \\
&+ E\left[\frac{1}{CPR}\right] \times \int_0^{2\pi}\int_0^{\pi} \exp(-jD_{mn}(t)) \\
&\times F_m^{Tx(v)}(\vartheta,\varphi) F_n^{Tx(v)}(\vartheta,\varphi) p^{Tx}(\vartheta,\varphi) sin\vartheta d\vartheta d\varphi \\
&\times \int_0^{2\pi}\int_0^{\pi} \exp(-jD_{pq}(t)) \\
&\times F_p^{Rx(v)}(\theta,\phi) F_q^{Rx(v)}(\theta,\phi) p^{Rx}(\theta,\phi) sin\theta d\theta d\phi)
\end{aligned}$$

(23)

where $P_{pm}$ and $P_{qn}$ are the power transformed in the subchannels $Tx_m - Rx_p$ and $Tx_n - Rx_q$, respectively.

Now for channel realization purpose, Monte Carlo simulation method will be used to compute the channel capacity. After constructing the spatial correlation matrix $\mathbf{R}$, whose elements are obtained by (11), and considering $\mathbf{H}_w$ as a matrix with complex Gaussian elements with $US \times 1$ dimension, channel matrix instances can be calculated by

$$\mathbf{H} = \text{unvec}\left(\mathbf{R}^{1/2}\text{vec}(\mathbf{H}_w)\right) \quad (24)$$

It is assumed that the MIMO system has no channel state information (CSI) and only the receiver knows the channel realization. This evinces that the power is equally distributed among the transmitter antennas and signals are independent. Under this condition, the channel capacity of the MIMO channel can be expressed as

$$\mathbf{C} = \log\left[\det\left(\mathbf{I} + \frac{\rho}{S}\mathbf{HH}^T\right)\right] \quad (25)$$

where $\mathbf{I}$ is the identity matrix, $\rho$ is the average SNR and $\mathbf{H}$ is the normalized channel matrix calculated by (9).
Considering power constraint and power dissipation at subchannels of multi-polarized MIMO systems, the SNR $\rho$ can be written as

$$\rho = \rho_0 / (1 + E[1/\text{XPD}]) \quad (26)$$

As it can be seen in (23), the channel capacity is dependent on the channel matrix $\mathbf{H}$, so the instantaneous channel capacity is a random value. In this paper, for channel realization purpose, we produced 10000 instances and applied Monte Carlo simulation method to achieve the realized instantaneous channel capacity.

## V. NUMERICAL RESULTS AND DISCUSSIONS

The above 4D channel model encompasses transmitter and receiver antennas properties such as antenna spacing, orientation, polarization and pattern as well as statistical environmental characteristics such as depolarizations, angle of departure (AoD), angle of arrival (AoA), and moving scatterers. In the numerical results, the effect of deterministically and stochastically moving scatterers on the performance of polarized MIMO systems from the perspective of channel correlation and capacity will be discussed. For this purpose, a $2 \times 2$ MIMO system with half a wavelength antenna elements configuration is considered. To derive the parameters associated with the mixture of VMF distribution of an environment, data processing using Soft-Expectation Maximization (Soft-EM) Algorithm on experimental data of that environment is needed. For simulation purpose, the result of mixture of VMF parameters in [41] considering 10 clusters of scatterers in the receiver side is used to obtain the AoA profile. For the transmitter side, one cluster of scatterers is assumed with the mean elevation and azimuth angles of $(\vartheta_0, \varphi_0) = (90°, 0°)$ and concentric parameter of $\kappa = 10$. As shown in Fig.2, the maximum incoming wave power happens at $(\theta, \phi) = (90°, 330°)$. It is assumed that the mean $\text{XPD}_v = \text{XPD}_h = 9\text{dB}$, the mean $\text{CPR} = 2\text{dB}$ and the reference SNR is $\rho_0 = 10\text{dB}$. The transmit and receive antenna array centers are set to be at the origin of the coordinate system and

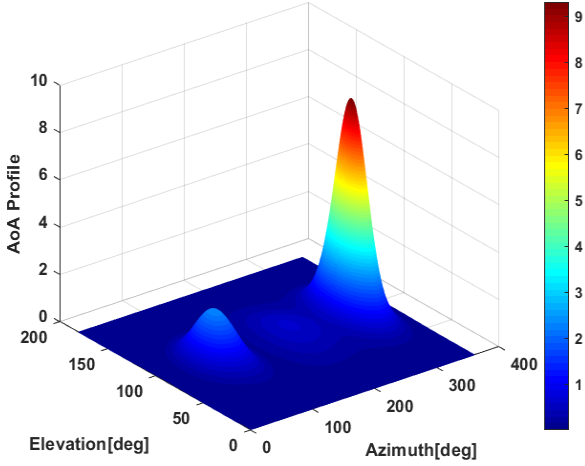

Fig.2. Angle of Arrival (AoA) due to 10 clusters of mixture of VMF distribution (at t=0 sec).

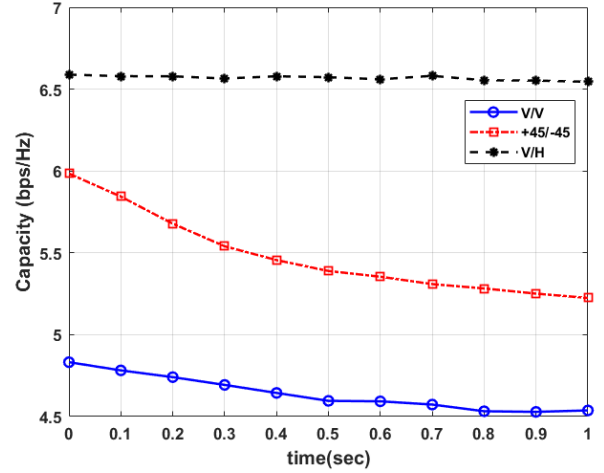

Fig.4. Time domain channel capacity with moving scatterers.

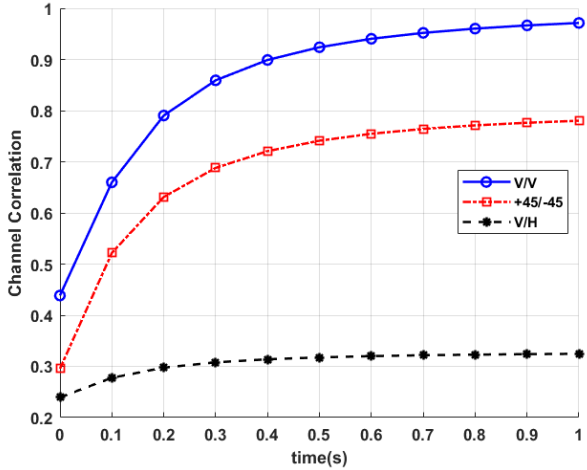

Fig.3. Time domain channel correlation considering moving scatterers

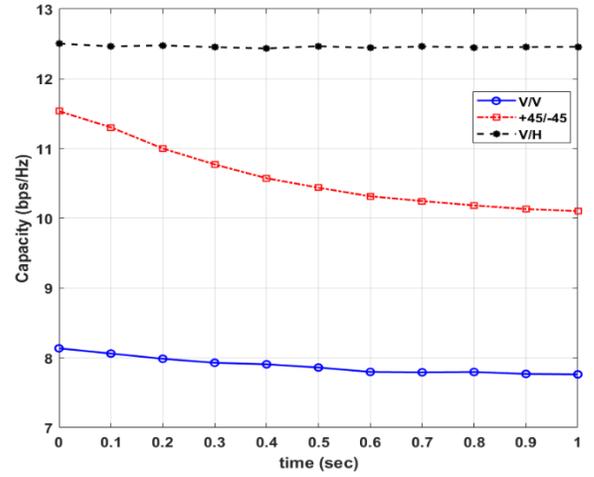

Fig.5. Time domain channel capacity considering moving scatterers (SNR=20dB).

$(x_{Rx}, y_{Rx}, z_{Rx}) = (0,100\text{m},0)$, respectively and $\delta_T = \delta_R = 0.1\lambda$ where $\lambda$ is the wavelength. Firstly, we will analyze the behavior of channel statistics due to radially moving scatterers as a deterministic motion then we will set $v^{Tx} = v^{Rx} = 0$m/s to observe the effect of BM random motion of clusters on the surface of the spheres. For the deterministic motion part, the initial assumed radius of sphers for scatterers at time t=0 sec in the transmitter and receiver side is $R^{Tx} = R^{Rx} = 1$m and the clusters of scatterers are considered to radially moving far away from their antenna array center with the velocity of $v^{Tx} = v^{Rx} = 10$m/s. Figure.3 illustrates the channel correlation behavior for this case for three different polarizations as Vertical-Vertical (V/V), slanted $\pm 45^0$ and Vertical-Horizontal (V/H) polarizations. It is found that while scatterers move away from their antenna array center by time, the channel correlation increases for all polarization sets. This indicates that specially for the receiver case, scatterers play a significant role to guide incoming waves to the antennas and the further the scatterers move from the antenna elements the higher channel correlation is expected. As it can be seen, the V/V polarization has the highest correlation, while the V/H polarization has the lowest and the slanted $\pm 45^0$ polarization shows a moderate performance so it is expected to experience higher capacity for V/H polarization. Next, Fig.4 represents the corresponding channel capacity behavior and as expected, generally the capacity gradually decreases over time since the correlation enlarges in the previous mode. Here, while V/V polarization has the lowest capacity V/H polarization is about 6.6 bps/Hz, or 41% higher than V/V configuration. Also, V/H polarization behaves almost independent of time which indicates that for rich scattering environments with many moving scatterers such as urban areas, the V/H polarization is a better candidate to use and will provide a greater and more stable capacity. Furthermore, Fig.5 is for the case where SNR is set to be 20dB to see the effect of SNR on the capacity of a channel in the presence of moving scatterers. As it can be seen, by increasing SNR all polarizations illustrate higher capacity compared to the previous scenario and the $\pm 45^0$ and V/V polarizations experience higher change while the V/H polarization maintains its independent behavior for moving scatterers which again

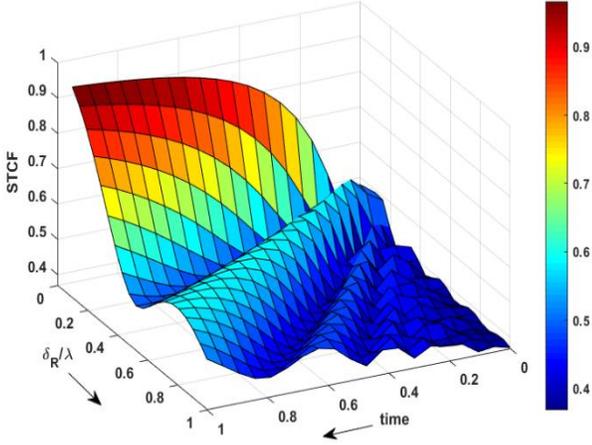

Fig.6. Space Time Correlation Function associated with moving scatterers.

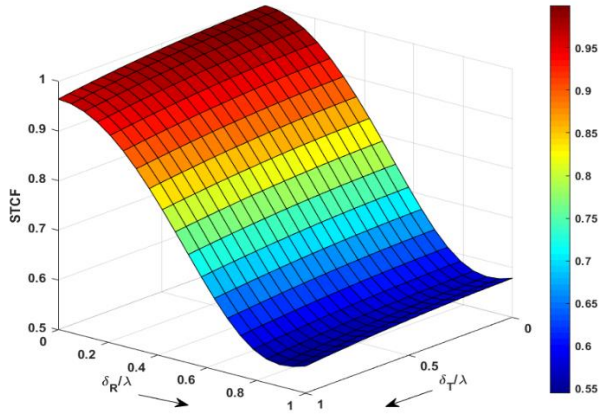

Fig.7. Space Time Correlation Function associated with different antenna element spacings at $t=2\text{sec}$.

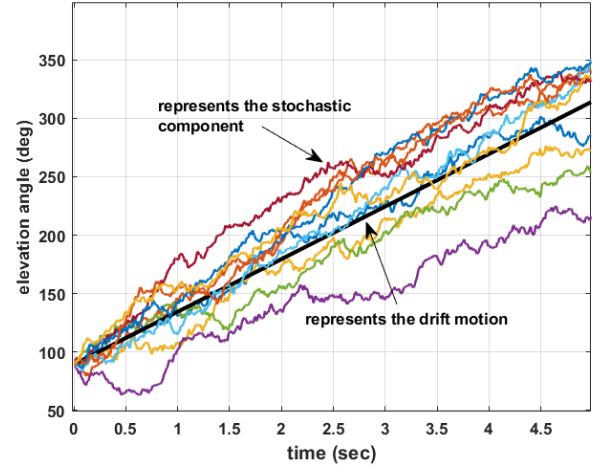

Fig.8. Brownian Motions with drift for a start point of $\theta_s = 90^0$ (note that the black motion path is for the deterministic motion as it represents just the drift motion)

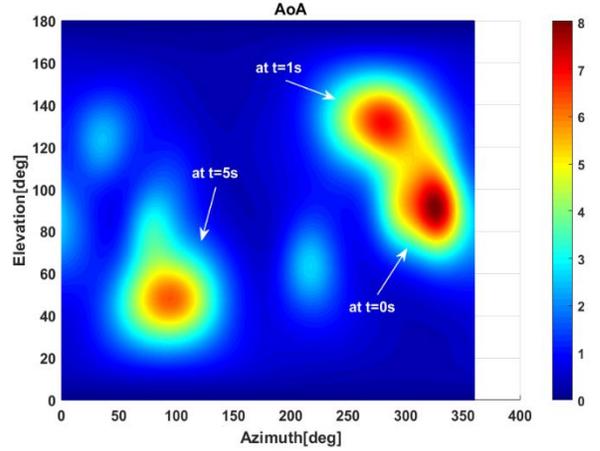

Fig.9. Contour plot of AoA for three time instants (t=0sec, 1sec and 5sec)

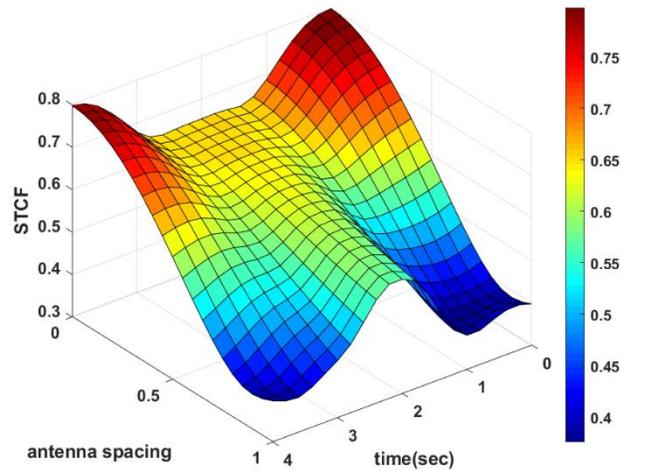

Fig.10. Space Time Correlation Function for stochastically moving scatterers.

implies its advantage from the perspective of both high and stable capacity. In fact, both Fig.4 and Fig.5 indicate that the dual polarized configurations overcome the single polarized arrangement and this result becomes even more distinct at high SNR regime. Figure.6 illustrates the dependency of STCF on both time and receiver antenna element spacing. This indicates the non-stationary nature of STFC for the moving scatterers and displays that the correlation decreases as the receiver spacing increases. Hence, this property in MIMO systems can be used to increase channel capacity depending on the scatterers distribution of the intended environment. Figure.7 demonstrates a comparison between the effect of both transmitter and receiver spacing ($\delta_T$ and $\delta_R$) on STCF. As it can be seen, while the STCF decreases by increasing the antenna spacings, it diminishes dramatically along the receiver spacing. Therefore, to gain a low correlation and high capacity, a larger antenna spacing is needed at receiver side, however a small antenna spacing at transmitter side satisfies the antenna gain diversity. Next, we will discuss the effect of randomly moving clusters which are supposed to move based on the proposed Brownian Motion path model on the surface of the spheres. The Brownian Motion for 10 trajectory paths are demonstrated in Fig.8. For this case, we set the motion parameters as start point of $\theta_s = 90\,\text{deg}$, drift parameter of $\omega_\theta^d = 45\,\text{deg/sec}$ and

$\sigma_\theta = 2\deg$. As it can be seen from the figure, since in a dynamic system we can have infinite BM paths for a single cluster, we need to perform channel realization using Monte Carlo Simulations this time due to the randomness of BMs. Hence, the clusters at receiver side are considered to have BM trajectories with the path motion characteristics of $(\omega_\theta^d, \omega_\phi^d) = (45, -45)\deg/\sec$ and $\sigma_\theta = \sigma_\phi = 2\deg$. The contour plot of angle of arrival for three different snapshots of t=0sec, 1sec and 5sec is provided by Fig.9 which is the result of applying these BMs on the initial AoA shown in Fig.2. Since the maximum power at t=0s occurs at $(\theta, \phi) = (90^0, 330^0)$ the maximum AoA amounts at t=1sec and t=5sec are expected to happen at around $(\theta, \phi) = (135^0, 285^0)$ and $(\theta, \phi) = (45^0, 105^0)$, respectively which matches well with the result. An interesting observation is that the maximum amplitudes of AoA decrease over time as the clusters experience random motions (considering $\sigma$) and thus the power distributes and degrades over time which this fact is confirmed by this result. Finally, Fig.10 represents the correlation over time and antenna spacing. For this case, motion parameters are set to be $(\omega_\theta^d, \omega_\phi^d) = (45, -15)\deg/\sec$ and $\sigma_\theta = \sigma_\phi = 2\deg$. The channel correlation is symmetric over the moment t=2sec which is specially because of the effect of elevation angle speed on the AoA over time.

## VI. CONCLUSION

MIMO systems are expected to play a critical role in high-data-rate transformation in 6G technology applications, and they will be in high demand in the near future for emerging scenarios of next-generation communications. As a result, their channel models are always expanding to include various components of cutting-edge technologies. In this paper, we investigated a 4D channel modeling for polarized MIMO systems with moving clusters of scatterers and arrive at a closed form formula. A novel motion path model inspired from the Brownian Motions was proposed which is capable of analyzing randomly moving objects in a dynamic system. The scatterers are assumed to have a mixture of VMF distributions, and Monte Carlo simulations are conducted for channel realization purpose. The results of the statistical channel characteristics demonstrate that the effect of moving scatterers should not be overlooked. Based on the results, moving scatterers have different effect on the performance of various polarization configurations. Also, the dual polarized configurations outperform the single polarized arrangements; this behavior is even more distinct in high SNR regimes. It is demonstrated that the antenna spacing contributes significantly in the behavior of STCF and specially the receiver antenna spacing has a notable impact. Therefore, to obtain a higher channel capacity, larger receiver antenna spacing is required while a small transmitter antenna spacing is sufficient for gain diversity. The agreement between the behavior of our channel correlation results and that of the literature [37] and [49-51] confirms the verification of our model. The procedure applied in this work is for linear array antennas, however it can be extended to any kind of arrays. Because of the versatility of the proposed model, it can be used to investigate the effect of moving scatterers in a variety of dynamic systems, including V2V, A2G, and A2A communications.

## I. ACKNOWLEDGEMENTS

This research is supported by the KMW Inc. Hwaseong, South Korea, (Project No. G01200434).

REFERENCES

[1] Wang, Liang, et al. "Joint trajectory and passive beamforming design for intelligent reflecting surface-aided UAV communications: A deep reinforcement learning approach." *arXiv preprint arXiv:2007.08380* (2020).
[2] Yan, Chaoxing, et al. "A comprehensive survey on UAV communication channel modeling." *IEEE Access* 7 (2019): 107769-107792.
[3] Ma, Zhangfeng, et al. "A Non-Stationary Geometry-Based MIMO Channel Model for Millimeter-Wave UAV Networks." *IEEE Journal on Selected Areas in Communications* (2021).
[4] Shafi, Mansoor, et al. "Polarized MIMO channels in 3-D: models, measurements and mutual information." *IEEE Journal on Selected Areas in Communications* 24.3 (2006): 514-527.
[5] Dao, Manh-Tuan, et al. "3D polarized channel modeling and performance comparison of MIMO antenna configurations with different polarizations." *IEEE Transactions on Antennas and Propagation* 59.7 (2011): 2672-2682.
[6] Ng, Kah Heng, et al. "Efficient multielement ray tracing with site-specific comparisons using measured MIMO channel data." *IEEE Transactions on Vehicular Technology* 56.3 (2007): 1019-1032.
[7] Shi, Zhiyuan, et al. "Modeling of wireless channel between UAV and vessel using the FDTD method." *10th International Conference on Wireless Communications, Networking and Mobile Computing (WiCOM 2014)*. IET, 2014.
[8] Kazemi, Mohammad Jafar, Abdolali Abdipur, and Abbas Mohammadi. "Indoor propagation MIMO channel modeling in 60 GHz using SBR based 3D ray tracing technique." *2012 Second Conference on Millimeter-Wave and Terahertz Technologies (MMWaTT)*. IEEE, 2012.
[9] Stabler, Oliver, and Reiner Hoppe. "MIMO channel capacity computed with 3D ray tracing model." *2009 3rd European Conference on Antennas and Propagation*. IEEE, 2009.
[10] He, Danping, et al. "The design and applications of high-performance ray-tracing simulation platform for 5G and beyond wireless communications: A tutorial." *IEEE Communications Surveys & Tutorials* 21.1 (2018): 10-27.
[11] Karedal, Johan, et al. "A geometry-based stochastic MIMO model for vehicle-to-vehicle communications." *IEEE transactions on wireless communications* 8.7 (2009): 3646-3657.
[12] Cheng, Xiang, et al. "An adaptive geometry-based stochastic model for non-isotropic MIMO mobile-to-mobile channels." *IEEE transactions on wireless communications* 8.9 (2009): 4824-4835.
[13] Zeng, Linzhou, et al. "A 3D geometry-based stochastic channel model for UAV-MIMO channels." *2017 IEEE Wireless Communications and Networking Conference (WCNC)*. IEEE, 2017.
[14] Cheng, Xiang, and Yiran Li. "A 3-D geometry-based stochastic model for UAV-MIMO wideband nonstationary channels." *IEEE Internet of Things Journal* 6.2 (2018): 1654-1662.
[15] Chang, Hengtai, et al. "A 3D non-stationary wideband GBSM for low-altitude UAV-to-ground V2V MIMO channels." *IEEE Access* 7 (2019): 70719-70732.
[16] Ghazal, Ammar, et al. "A nonstationary wideband MIMO channel model for high-mobility intelligent transportation systems." *IEEE Transactions on Intelligent Transportation Systems* 16.2 (2014): 885-897.
[17] Jiang, Hao, et al. "A 3-D non-stationary wideband geometry-based channel model for MIMO vehicle-to-vehicle communications in tunnel environments." *IEEE Transactions on Vehicular Technology* 68.7 (2019): 6257-6271.
[18] Acosta-Marum, Guillermo, and Mary Ann Ingram. "Six time-and frequency-selective empirical channel models for vehicular wireless LANs." *IEEE Vehicular Technology Magazine* 2.4 (2007): 4-11.
[19] Matolak, David W., and Ruoyu Sun. "Air–ground channel characterization for unmanned aircraft systems—Part I: Methods, measurements, and models for over-water settings." *IEEE Transactions on Vehicular Technology* 66.1 (2016): 26-44.

[20] Sun, Ruoyu, and David W. Matolak. "Air–ground channel characterization for unmanned aircraft systems part II: Hilly and mountainous settings." *IEEE Transactions on Vehicular Technology* 66.3 (2016): 1913-1925.

[21] Matolak, David W., and Ruoyu Sun. "Air–ground channel characterization for unmanned aircraft systems—Part III: The suburban and near-urban environments." *IEEE Transactions on Vehicular Technology* 66.8 (2017): 6607-6618.

[22] Zajic, Alenka G., and Gordon L. Stuber. "Three-dimensional modeling and simulation of wideband MIMO mobile-to-mobile channels." *IEEE Transactions on Wireless Communications* 8.3 (2009): 1260-1275.

[23] Jiang, Hao, Zaichen Zhang, and Guan Gui. "Three-dimensional non-stationary wideband geometry-based UAV channel model for A2G communication environments." *IEEE Access* 7 (2019): 26116-26122.

[24] Patzold, Matthias, Bjorn Olav Hogstad, and Neji Youssef. "Modeling, analysis, and simulation of MIMO mobile-to-mobile fading channels." *IEEE Transactions on Wireless Communications* 7.2 (2008): 510-520.

[25] Zajic, Alenka G., and Gordon L. Stuber. "Three-dimensional modeling, simulation, and capacity analysis of space–time correlated mobile-to-mobile channels." *IEEE Transactions on Vehicular Technology* 57.4 (2008): 2042-2054.

[26] Paier, Alexander, et al. "Characterization of vehicle-to-vehicle radio channels from measurements at 5.2 GHz." *Wireless personal communications* 50.1 (2009): 19-32.

[27] Gehring, Andreas, et al. "Empirical channel stationarity in urban environments." *Proceedings of the 4th European Personal Mobile Communications Conference (EPMCC'01)*. 2001.

[28] Umansky, Dmitry, and Matthias Patzold. "Stationarity test for wireless communication channels." *GLOBECOM 2009-2009 IEEE Global Telecommunications Conference*. IEEE, 2009.

[29] Ispas, Adrian, et al. "Analysis of local quasi-stationarity regions in an urban macrocell scenario." *2010 IEEE 71st Vehicular Technology Conference*. IEEE, 2010.

[30] Yuan, Yi, et al. "Novel 3D geometry-based stochastic models for non-isotropic MIMO vehicle-to-vehicle channels." *IEEE Transactions on Wireless Communications* 13.1 (2013): 298-309.

[31] Sun, Ruoyu, and David W. Matolak. "Air–ground channel characterization for unmanned aircraft systems part II: Hilly and mountainous settings." *IEEE Transactions on Vehicular Technology* 66.3 (2016): 1913-1925.

[32] Borhani, Alireza, and Matthias Pätzold. "Modelling of non-stationary mobile radio channels using two-dimensional Brownian motion processes." *2013 International Conference on Advanced Technologies for Communications (ATC 2013)*. IEEE, 2013.

[33] Borhani, Alireza, and Matthias Pätzold. "A novel non-stationary channel model utilizing Brownian random paths." *REV Journal on Electronics and Communications* 4.1-2 (2014).

[34] Borhani, Alireza, Gordon L. Stüber, and Matthias Pätzold. "A random trajectory approach for the development of nonstationary channel models capturing different scales of fading." *IEEE Transactions on Vehicular Technology* 66.1 (2016): 2-14.

[35] Borhani, Alireza, and Matthias P̈atzold. "Modeling of vehicle-to-vehicle channels in the presence of moving scatterers." *2012 IEEE Vehicular Technology Conference (VTC Fall)*. IEEE, 2012.

[36] Borhani, Alireza, and Matthias Pätzold. "Correlation and spectral properties of vehicle-to-vehicle channels in the presence of moving scatterers." *IEEE Transactions on Vehicular Technology* 62.9 (2013): 4228-4239.

[37] Jiang, Hao, et al. "A non-stationary geometry-based scattering vehicle-to-vehicle MIMO channel model." *IEEE communications letters* 22.7 (2018): 1510-1513.

[38] Zajić, Alenka G. "Impact of moving scatterers on vehicle-to-vehicle narrow-band channel characteristics." *IEEE Transactions on Vehicular Technology* 63.7 (2014): 3094-3106.

[39] Zhao, Xiongwen, et al. "Two-cylinder and multi-ring GBSSM for realizing and modeling of vehicle-to-vehicle wideband MIMO channels." *IEEE Transactions on Intelligent Transportation Systems* 17.10 (2016): 2787-2799.

[40] Ma, Zhangfeng, et al. "A Non-Stationary Geometry-Based MIMO Channel Model for Millimeter-Wave UAV Networks." *IEEE Journal on Selected Areas in Communications* (2021).

[41] Mammasis, Konstantinos, Robert W. Stewart, and John S. Thompson. "Spatial fading correlation model using mixtures of Von Mises Fisher distributions." *IEEE Transactions on Wireless Communications* 8.4 (2009): 2046-2055.

[42] Einstein, Albert. "Über die von der molekularkinetischen Theorie der Wärme geforderte Bewegung von in ruhenden Flüssigkeiten suspendierten Teilchen." *Annalen der physik* 4 (1905).

[43] Beichelt, Frank. *Stochastic processes in science, engineering and finance*. Chapman and Hall/CRC, 2006.

[44] Pitman, Jim, and Marc Yor. "A guide to Brownian motion and related stochastic processes." *arXiv preprint arXiv:1802.09679* (2018).

[45] Mubayi, Anuj, et al. "Studying complexity and risk through stochastic population dynamics: Persistence, resonance, and extinction in ecosystems." *Handbook of Statistics*. Vol. 40. Elsevier, 2019. 157-193.

[46] Borhani, Alireza, and Matthias Pätzold. "A highly flexible trajectory model based on the primitives of Brownian fields—Part I: Fundamental principles and implementation aspects." *IEEE Transactions on Wireless Communications* 14.2 (2014): 770-780.

[47] Marković, Ivan, et al. "Direction-only tracking of moving objects on the unit sphere via probabilistic data association." *17th International Conference on Information Fusion (FUSION)*. IEEE, 2014.

[48] Chiuso, Alessandro, and Giorgio Picci. "Visual tracking of points as estimation on the unit sphere." *The confluence of vision and control*. Springer, London, 1998. 90-105.

[49] Chelli, Ali, and Matthias Patzold. "The impact of fixed and moving scatterers on the statistics of MIMO vehicle-to-vehicle channels." *VTC Spring 2009-IEEE 69th Vehicular Technology Conference*. IEEE, 2009.

[50] Zhu, Qiuming, et al. "3D non-stationary geometry-based multi-input multi-output channel model for UAV-ground communication systems." *IET Microwaves, Antennas & Propagation* 13.8 (2019): 1104-1112.

[51] Zhu, Qiuming, et al. "A novel 3D non-stationary UAV-MIMO channel model and its statistical properties." *China Communications* 15.12 (2018): 147-158.